\documentclass[a4paper,fleqn,usenatbib]{mnras}
 \pdfoutput=1
\usepackage{newtxtext,newtxmath}
\usepackage[T1]{fontenc}
\usepackage{ae,aecompl}

\usepackage{graphicx}	% Including figure files
\usepackage{amsmath}	% Advanced maths commands
\usepackage{amssymb}	% Extra maths symbols
\usepackage{ulem}       % strikethrough function
\usepackage[dvipsnames]{xcolor}

\newcommand{\Rp}{R_p}
\newcommand{\Rs}{R_{\star}}
\newcommand{\RpRs}{\Rp/\Rs}

\newcommand{\aRs}{a/\Rs}
\newcommand{\Tmid}{T_{\textnormal{mid}}}

\newcommand{\um}{\mu\textnormal{m}}

\newcommand{\kir}{\kappa_\textnormal{IR}}
\newcommand{\Cabun}{[\textnormal{C}/\textnormal{H}]}
\newcommand{\Oabun}{[\textnormal{O}/\textnormal{H}]}
\newcommand{\Z}{[\textnormal{M}/\textnormal{H}]}

\title[Water emission in the dayside spectrum of WASP-121b]{Confirmation of water emission in the dayside spectrum of the ultrahot Jupiter WASP-121b}

\author[T.\ Mikal-Evans et al.]{
Thomas Mikal-Evans,$^{1}$\thanks{E-mail: tmevans@mit.edu} David K.\ Sing,$^{2,3}$ Tiffany Kataria,$^{4}$ Hannah R.\ Wakeford,$^{5}$
\newauthor Nathan J.\ Mayne,$^{6}$ Nikole K.\ Lewis,$^{7}$ Joanna K.\ Barstow,$^{8}$ Jessica J.\ Spake$^{2}$ 
\\
$^{1}$Kavli Institute for Astrophysics and Space Research, Massachusetts Institute of Technology, 77 Massachusetts Avenue, \\
37-241, Cambridge, MA 02139, USA\\
$^{2}$Department of Earth \& Planetary Sciences, Johns Hopkins University, Baltimore, MD, USA\\
$^{3}$Department of Physics \& Astronomy, Johns Hopkins University, Baltimore, MD, USA\\
$^{4}$NASA Jet Propulsion Laboratory, 4800 Oak Grove Drive, Pasadena, CA 91109, USA\\
$^{5}$School of Physics, University of Bristol, HH Wills Physics Laboratory, Tyndall Avenue, Bristol BS8 1TL, UK\\
$^{6}$Astrophysics Group, University of Exeter, Exeter, EX4 2QL, UK\\
$^{7}$Department of Astronomy and Carl Sagan Institute, Cornell University, 122 Sciences Drive, 14853, Ithaca, NY, USA\\
$^{8}$Department of Physics and Astronomy, University College London, Gower Street, London WC1E 6BT, UK\\
\\
}

\date{Accepted 2020 May 17. Received 2020 April 30; in original form 2020 February 28.}

\pubyear{2020}

\begin{document}
\label{firstpage}
\pagerange{\pageref{firstpage}--\pageref{lastpage}}
\maketitle

\begin{abstract}
We present four new secondary eclipse observations for the ultrahot Jupiter WASP-121b acquired using the \textit{Hubble Space Telescope} Wide Field Camera 3. The eclipse depth is measured to a median precision of 60\,ppm across 28 spectroscopic channels spanning the $1.12$-$1.64\,\um$ wavelength range. This is a considerable improvement to the 90\,ppm precision we achieved previously for a single eclipse observation using the same observing setup. Combining these data with those reported at other wavelengths, a blackbody spectrum for WASP-121b is ruled out at $>6\sigma$ confidence and we confirm the interpretation of previous retrieval analyses that found the data is best explained by a dayside thermal inversion. The updated spectrum clearly resolves the water emission band at $1.3$-$1.6\,\um$, with higher signal-to-noise than before. It also fails to reproduce a bump in the spectrum at $1.25\,\um$ derived from the first eclipse observation, which had tentatively been attributed to VO emission. We conclude the latter was either a statistical fluctuation or a systematic artefact specific to the first eclipse dataset.
\end{abstract}

\section{Introduction}

In the case of a synchronously-orbiting, highly-irradiated planet, if the opacity of the atmosphere is lower at optical wavelengths than at infrared wavelengths, the vertical temperature profile of the dayside hemisphere is expected to decrease with decreasing pressure close to the infrared photosphere. This is essentially because most of the incident stellar radiation is at optical wavelengths. Therefore, if the optical opacity of the atmosphere is lower than the infrared opacity, the stellar radiation will be primarily deposited below the infrared photosphere, heating the atmosphere at those higher pressures. Conversely, if the optical opacity is higher than the infrared opacity, most of the heating by the host star will occur above the infrared photosphere, resulting in a thermal inversion. As such, thermal inversions are valuable diagnostics of the radiative processes at play in a planetary atmosphere, and in particular, the relative strength of absorption at optical versus infrared wavelengths.

Observationally, thermal inversions can be inferred by measuring the planetary emission spectrum and detecting opacity bands as emission rather than absorption features. The first detection of a spectrally-resolved emission feature for an exoplanet was made by \cite{2017Natur.548...58E} for WASP-121b, an ultrahot Jupiter discovered by \cite{2016MNRAS.tmp..312D}. This was done by observing a secondary eclipse with the \textit{Hubble Space Telescope} (HST) Wide Field Imaging Camera 3 (WFC3). The resulting dayside spectrum derived from these data revealed an H$_2$O emission band spanning the $\sim 1.3$-$1.6\,\um$ wavelength range, providing strong evidence for a dayside thermal inversion \citep{2017Natur.548...58E}. Additional observations made with the \textit{Spitzer Space Telescope} \citep{2020AJ....159..137G}, ground-based photometry \citep{2016MNRAS.tmp..312D,2019A&A...625A..80K}, HST \citep{2019MNRAS.488.2222M}, and the \textit{Transiting Exoplanet Survey Satellite} (TESS) \citep{2019arXiv190903010B,2019arXiv190903000D} have since extended the wavelength coverage of the WASP-121b emission spectrum considerably. In addition to the H$_2$O emission band, this combined dataset shows evidence for H$^{-}$ and CO emission, with retrieval analyses inferring a temperature profile that increases from $\sim 2500$\,K to $\sim 2800$\,K across the $\sim 30$ to $5$\,mbar pressure range \citep{2019MNRAS.488.2222M}.

It has not yet been possible to identify optical opacity source(s) in the dayside atmosphere of WASP-121b and definitively link them to the thermal inversion. Early studies focused on the strong optical absorbers TiO and VO as likely candidates for generating thermal inversions in highly-irradiated atmospheres, in which the temperatures are high enough ($\gtrsim 2000$\,K) for these species to be in the gas phase \citep{2003ApJ...594.1011H,2008ApJ...678.1419F}. Indeed, evidence for VO absorption has been uncovered in the transmission spectrum of WASP-121b, but not TiO \citep{2018AJ....156..283E}. However, recent theoretical work has highlighted that for ultrahot Jupiters with temperatures $\gtrsim 2700$\,K such as WASP-121b, much of the TiO and VO will likely be thermally dissociated on the dayside hemisphere, reducing their potency as thermal inversion drivers \citep{2018A&A...617A.110P,2018ApJ...866...27L}. Instead, for these hottest planets, thermal inversions may be generated by heavy metals in the gas phase, such as Fe and Mg, which have strong absorption lines in the near-ultraviolet and optical \citep{2018ApJ...866...27L}. Statistically significant detections of FeI, FeII, and MgII have been made in the transmission spectrum of WASP-121b \citep{2019AJ....158...91S,2020MNRAS.tmp..220G,2020arXiv200106836B,2020arXiv200107196C}, supporting this hypothesis. Other near-ultraviolet/optical absorbers such as NaH, MgH, FeH, SiO, AlO, and CaO have also been suggested \citep{2018ApJ...866...27L,2018A&A...617A.110P,2019MNRAS.485.5817G}, but no evidence has been uncovered for their presence in the atmosphere of WASP-121b to date.

This paper presents follow-up secondary eclipse observations for WASP-121b made with HST WFC3 that allow us to refine the dayside emission spectrum across the $1.12$-$1.64\,\um$ wavelength range. In Section \ref{sec:datared} we describe the observations and data reduction procedures, followed by the light curve fitting methodology in Section \ref{sec:lcfitting}. We discuss the results in Section \ref{sec:discussion} and give our conclusions in Section \ref{sec:conclusion}.

\section{Observations and data reduction} \label{sec:datared}

Two full-orbit phase curves of WASP-121b were observed with HST/WFC3 on 2018 March 12-13 and 2019 February 3-4 (G.O.\ 15134; P.I.s\ Mikal-Evans \& Kataria). For both visits, the target was observed for approximately 40.3 hours over 26 contiguous HST orbits. Each visit was scheduled to include two consecutive secondary eclipses. Here, we present an analysis of the four secondary eclipses acquired in this manner. The full phase curve will be presented in a future publication (Mikal-Evans et al., \textit{in prep}).

A similar observing setup to that used previously in \cite{2016ApJ...822L...4E,2017Natur.548...58E} was adopted. Observations for both visits used the G141 grism, which encompasses the $1.12$-$1.64\,\um$ wavelength range with a spectral resolving power of $R \sim 130$ at $\lambda = 1.4\,\um$. The forward spatial-scanning mode was used and only a $256 \times 256$ subarray was read out from the detector with the SPARS10 sampling sequence and 15 non-destructive reads per exposure ($\textnormal{NSAMP}=15$), corresponding to exposure times of 103\,s. The only difference between \cite{2016ApJ...822L...4E,2017Natur.548...58E} and the current observing setup was that a slower spatial scan rate of $0.073$\,arcsec\,s$^{-1}$ was used, compared to $0.120$\,arcsec\,s$^{-1}$ for the earlier observations. This resulted in shorter scans across approximately 60 pixel-rows of the cross-dispersion axis, leaving more space on the detector for background estimation.  With this setup, we obtained 15 exposures in the first HST orbit following acquisition and 16 exposures in each subsequent HST orbit. Typical peak frame counts were $\sim 37,000$ electrons per pixel for both visits, which is within the recommended range derived from an ensemble analysis of WFC3 spatial-scan data spanning eight years \citep{2019wfc..rept...12S}.

Spectra were extracted from the raw data frames using a custom-built Python pipeline, which has been described previously \citep{2016ApJ...822L...4E,2017Natur.548...58E,2019MNRAS.488.2222M} and is similar to others employed in the field \citep[e.g.][]{2014Natur.505...66K,2014Natur.505...69K,2017Sci...356..628W,2018AJ....155...29W,2018MNRAS.474.1705N}. For each exposure, we took the difference between successive non-destructive reads and applied a 50-pixel-wide top-hat filter along the cross-dispersion axis, before summing to produce final reconstructed images. The top-hat filter applied in this way has the effect of removing contamination from nearby sources and most cosmic ray strikes on the detector. The target spectrum was then extracted from each image by integrating the flux within a rectangular aperture spanning the full dispersion axis and 100 pixels along the cross-dispersion axis, centered on the central cross-dispersion row of the scan. Background fluxes were assumed to be wavelength-independent and subtracted from each spectrum. These were estimated by taking the median pixel count within a $10 \times 170$ pixel box located away from the target spectrum on the 2D reconstructed image, with typical background levels integrated over the full $103\,$s exposures starting at $\sim 150$\,electrons\,pixel$^{-1}$ and dropping to $\sim 110$\,electrons\,pixel$^{-1}$ over each HST orbit. The wavelength solution was determined by cross-correlating the final spectrum of each visit against a model stellar spectrum, as described in \cite{2016ApJ...822L...4E}.

\section{Light Curve Fitting} \label{sec:lcfitting}

% Made using the figures.TrEcWhiteLCFits() routine.
\begin{figure}
\centering  % this centres figure in column
\includegraphics[width=\columnwidth]{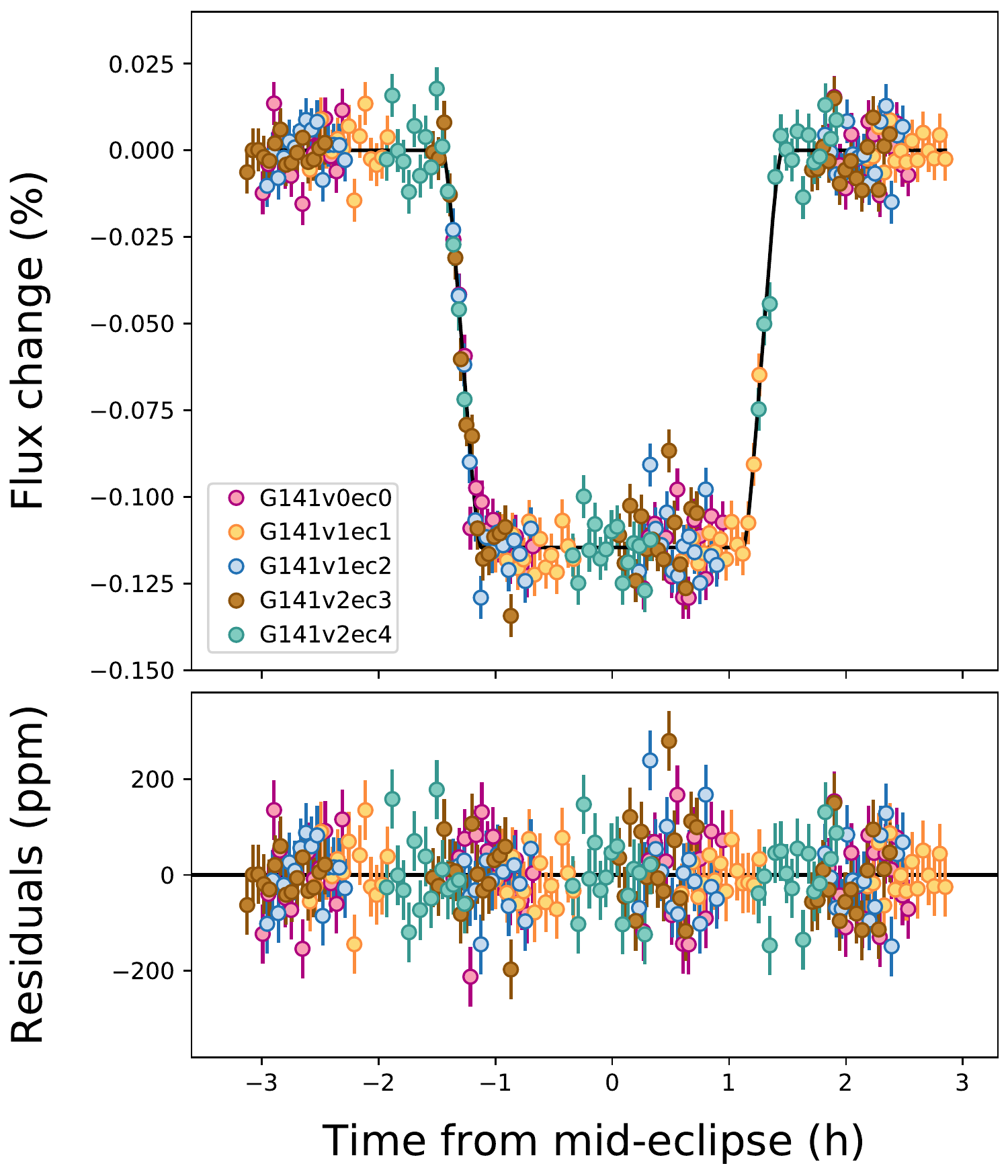}
\caption{\textit{(Top panel)} Phase-folded white light curve with eclipse model after removing systematics components of a joint fit to all five eclipses. \textit{(Bottom panel)} Model residuals.}
\label{fig:whitelc}
\end{figure}

White light curves were produced by summing the flux of each spectrum across the full wavelength range. We then fit the resulting light curves using the methodology described in \cite{2019MNRAS.488.2222M}, using a Gaussian process (GP) model to account for instrumental systematics. In addition to the four new eclipses presented in this study, we also analyzed an eclipse acquired as part of an earlier HST program (GO-14767; P.I.s\ Sing \& Lopez-Morales) that was originally published in \cite{2017Natur.548...58E}. We refer to the latter eclipse as G141v0ec0; the two eclipses observed as part of the 2018 visit as G141v1ec1 and G141v1ec2; and the two eclipses observed as part of the 2019 visit as G141v2ec3 and G141v2ec4. All five eclipses were fit simultaneously, with separate systematics models and eclipse mid-times ($\Tmid$) for each dataset, and a shared eclipse depth. We set the orbital period equal to 1.2749247646 day \citep{2019AJ....158...91S}, and the normalized semimajor axis ($\aRs$) and impact parameter ($b$) were fixed to the same values adopted in \cite{2019MNRAS.488.2222M}: namely, $\aRs=3.86$ and $b=0.06$. The resulting light curve fit is shown in Figure \ref{fig:whitelc} and the inferred eclipse parameters are reported in Table \ref{table:whitefit}. 

\begin{table}
\begin{minipage}{\columnwidth}
  \centering
%\scriptsize
\caption{MCMC results for the joint fit to all five eclipse white light curves. Quoted values are the posterior medians and uncertainties give the $\pm 34$\% credible intervals about the median. \label{table:whitefit}}
\begin{tabular}{cccc}
\hline \\ 
Parameter & Dataset & Value \medskip \\ \cline{1-3}
&&& \\
Eclipse depth (ppm)                        &    All    & $1150_{-19}^{+21}$ \\
  $\Tmid$ (MJD$_{\textnormal{UTC}}$) & G141v0ec0 & $2457703.45707_{-0.00081}^{+0.00077}$ \smallskip \\
                                 & G141v1ec1 & $2458190.47621_{-0.00115}^{+0.00142}$ \smallskip \\ 
                                 & G141v1ec2 & $2458191.75090_{-0.00055}^{+0.00049}$ \smallskip \\ 
                                 & G141v2ec3 & $2458518.13064_{-0.00073}^{+0.00073}$ \smallskip \\ 
                                 & G141v2ec4 & $2458519.40663_{-0.00056}^{+0.00058}$ \\ \\ \hline
\end{tabular}
\end{minipage}
\end{table}

% Figure 2. Example Spectroscopic Light Curve: same as Figure 1.
\begin{figure}
\centering  % this centres figure in column
\includegraphics[width=\columnwidth]{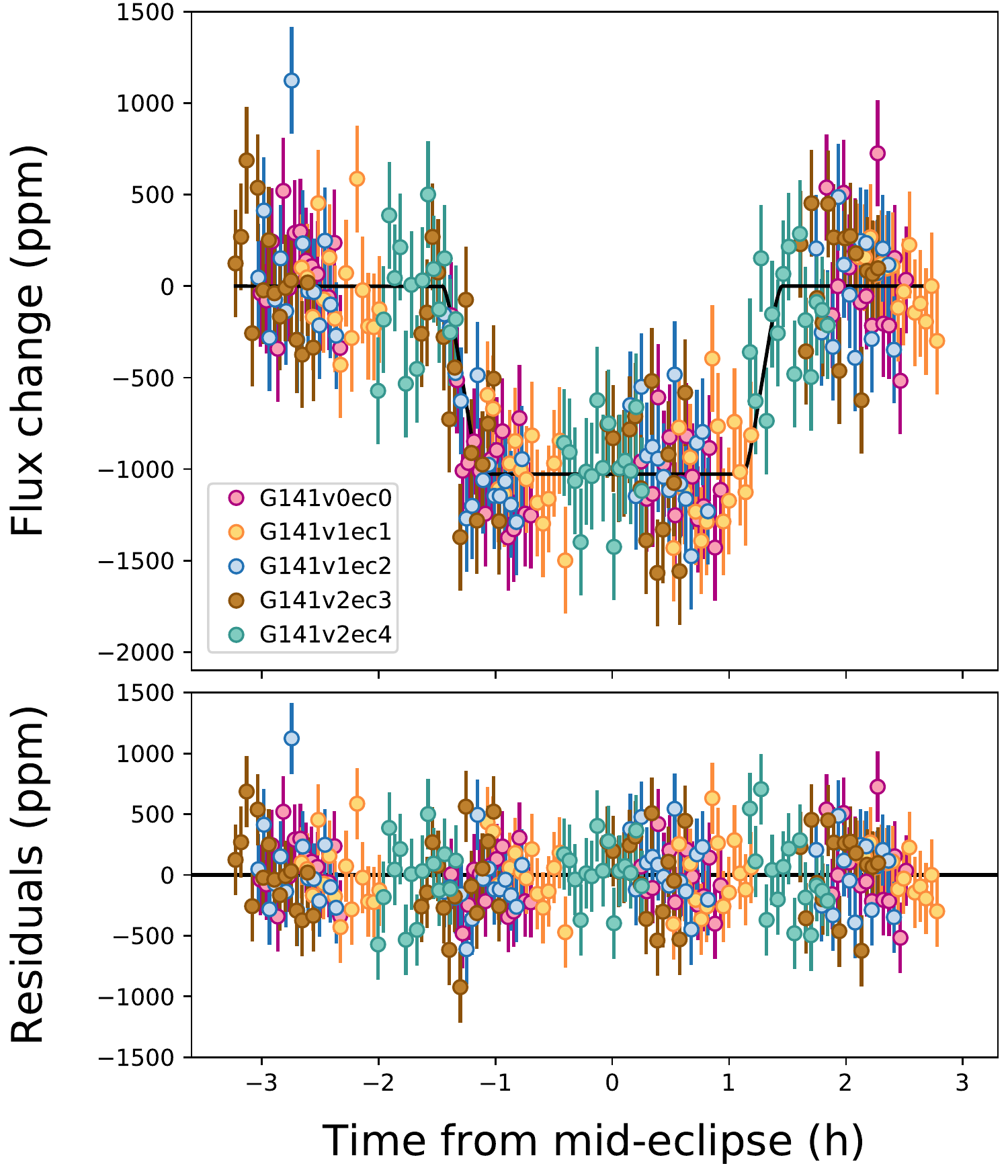}
\caption{The same as Figure \ref{fig:whitelc}, but for an example spectroscopic light curve spanning $1.231$-$1.249\,\um$ in wavelength.}
\label{fig:speclc}
\end{figure}

Next, we generated spectroscopic light curves in 28 wavelength channels, using the method described in \cite{2019MNRAS.488.2222M}. Each of these light curves was then fit with the same method used for the white light curve fit, but with the eclipse mid-times held fixed to the best-fit values obtained from the latter. An example light curve fit is shown in Figure \ref{fig:speclc}. For all datasets and spectroscopic light curves, the residual scatter was consistent with being photon noise limited. Inferred eclipse depths are reported in Table \ref{table:specfits}.

\begin{table}
  \begin{minipage}{\columnwidth}
  \centering
%\scriptsize
\caption{Eclipse depths inferred for each spectroscopic channel, quoted as median and $\pm 34$\% credible intervals from the MCMC fits. \label{table:specfits}}
\begin{tabular}{cc}
\hline \\ 
Wavelength ($\um$) & Eclipse depth (ppm) \medskip \\ \cline{1-2}
& \\
$1.120$-$1.138$ & $903_{-52}^{+53}$ \smallskip \\
$1.138$-$1.157$ & $991_{-60}^{+59}$ \smallskip \\
$1.157$-$1.175$ & $1002_{-56}^{+58}$ \smallskip \\
$1.175$-$1.194$ & $1029_{-49}^{+50}$ \smallskip \\
$1.194$-$1.212$ & $1066_{-58}^{+58}$ \smallskip \\
$1.212$-$1.231$ & $983_{-57}^{+54}$ \smallskip \\
$1.231$-$1.249$ & $1031_{-51}^{+48}$ \smallskip \\
$1.249$-$1.268$ & $1015_{-50}^{+55}$ \smallskip \\
$1.268$-$1.286$ & $994_{-48}^{+49}$ \smallskip \\
$1.286$-$1.305$ & $1028_{-53}^{+55}$ \smallskip \\
$1.305$-$1.323$ & $1008_{-60}^{+55}$ \smallskip \\
$1.323$-$1.342$ & $1077_{-56}^{+55}$ \smallskip \\
$1.342$-$1.360$ & $1160_{-52}^{+51}$ \smallskip \\
$1.360$-$1.379$ & $1110_{-57}^{+61}$ \smallskip \\
$1.379$-$1.397$ & $1262_{-59}^{+57}$ \smallskip \\
$1.397$-$1.416$ & $1360_{-56}^{+57}$ \smallskip \\
$1.416$-$1.434$ & $1193_{-56}^{+54}$ \smallskip \\
$1.434$-$1.453$ & $1304_{-54}^{+51}$ \smallskip \\
$1.453$-$1.471$ & $1331_{-63}^{+58}$ \smallskip \\
$1.471$-$1.490$ & $1342_{-57}^{+57}$ \smallskip \\
$1.490$-$1.508$ & $1304_{-60}^{+62}$ \smallskip \\
$1.508$-$1.527$ & $1276_{-62}^{+62}$ \smallskip \\
$1.527$-$1.545$ & $1210_{-65}^{+66}$ \smallskip \\
$1.545$-$1.564$ & $1307_{-60}^{+62}$ \smallskip \\
$1.564$-$1.582$ & $1388_{-66}^{+63}$ \smallskip \\
$1.582$-$1.601$ & $1299_{-69}^{+69}$ \smallskip \\
$1.601$-$1.619$ & $1270_{-63}^{+64}$ \smallskip \\
$1.619$-$1.638$ & $1286_{-68}^{+68}$ \\ \\ \hline
\end{tabular}
\end{minipage}
\end{table}

\section{Discussion} \label{sec:discussion}

% Figure 3. Updated emission spectrum.
\begin{figure*}
\centering  % this centres figure in column
\includegraphics[width=0.8\linewidth]{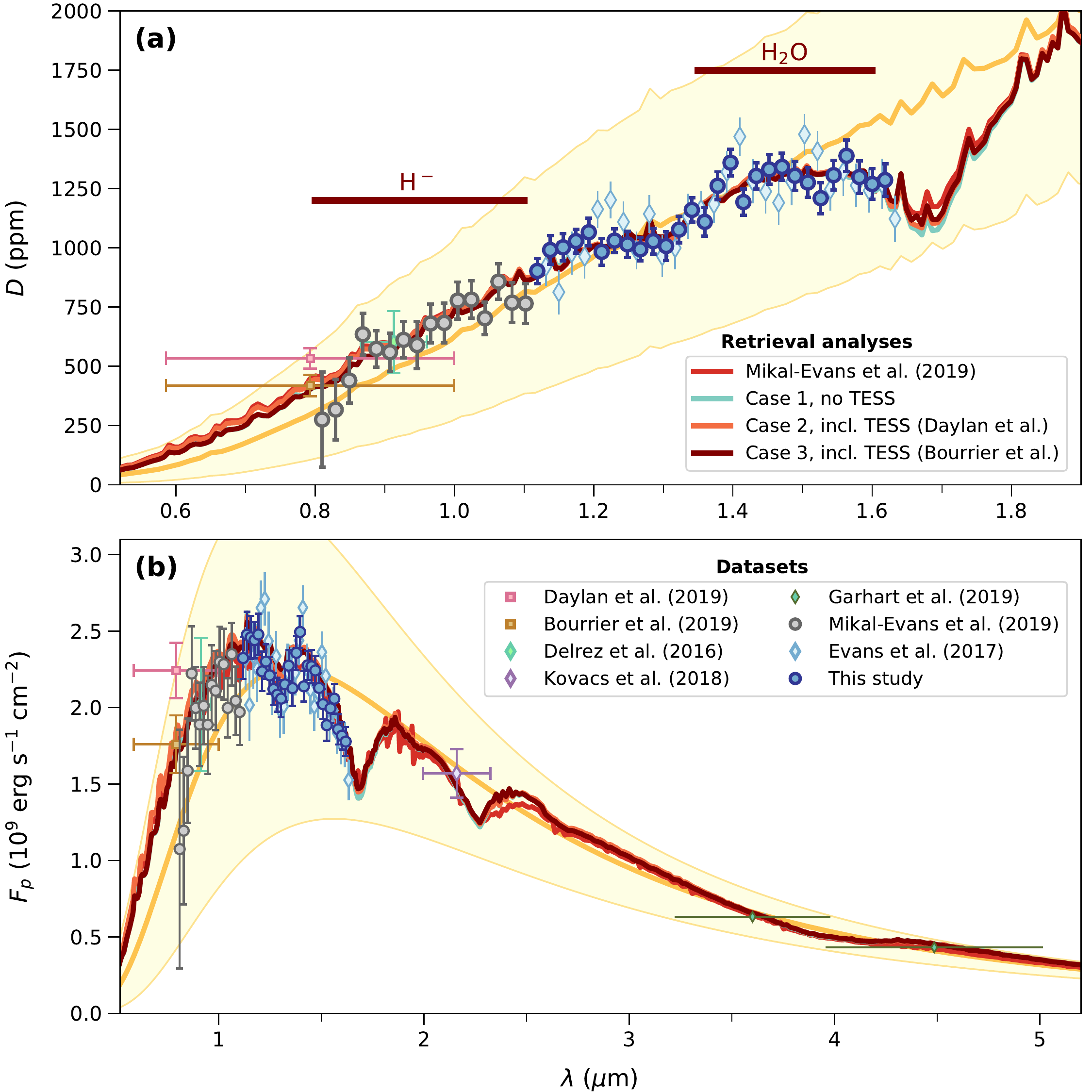}
\caption{\textit{(a)} Published eclipse depth measurements for WASP-121b across the red optical and near-infrared wavelength range covered by the TESS and HST WFC3 passbands. Note in particular the improved WFC3 G141 signal-to-noise achieved for the present study with five eclipse observations, compared to the original data presented in \citet{2017Natur.548...58E} for a single eclipse observation. \textit{(b)} Corresponding planetary emission extending out to longer wavelengths including the \textit{Spitzer} IRAC passbands. The errorbars for the IRAC measurements are not much larger than the marker symbols on this vertical scale. In both panels, the dark yellow line shows the spectrum assuming the planet radiates as a blackbody with a best-fit temperature of 2700\,K and the pale yellow envelope indicates blackbody spectra for temperatures of 2330\,K and 2970\,K. The latter encompass a plausible range of emission under limiting assumptions for the albedo and day-night heat recirculation. As labeled in panel \textit{(a)}, other solid lines show best-fit models obtained for the three retrieval analyses described in the main text, which all incorporate the updated WFC3 G141 spectrum, along with that obtained for our previous retrieval analysis published in \citet{2019MNRAS.488.2222M}. Spectral emission features due to H$^-$ and H$_2$O are also labeled in panel \textit{(a)}.
}
\label{fig:emspec}
\end{figure*}

\begin{figure*}
\centering  % this centres figure in column
\includegraphics[width=0.98\linewidth]{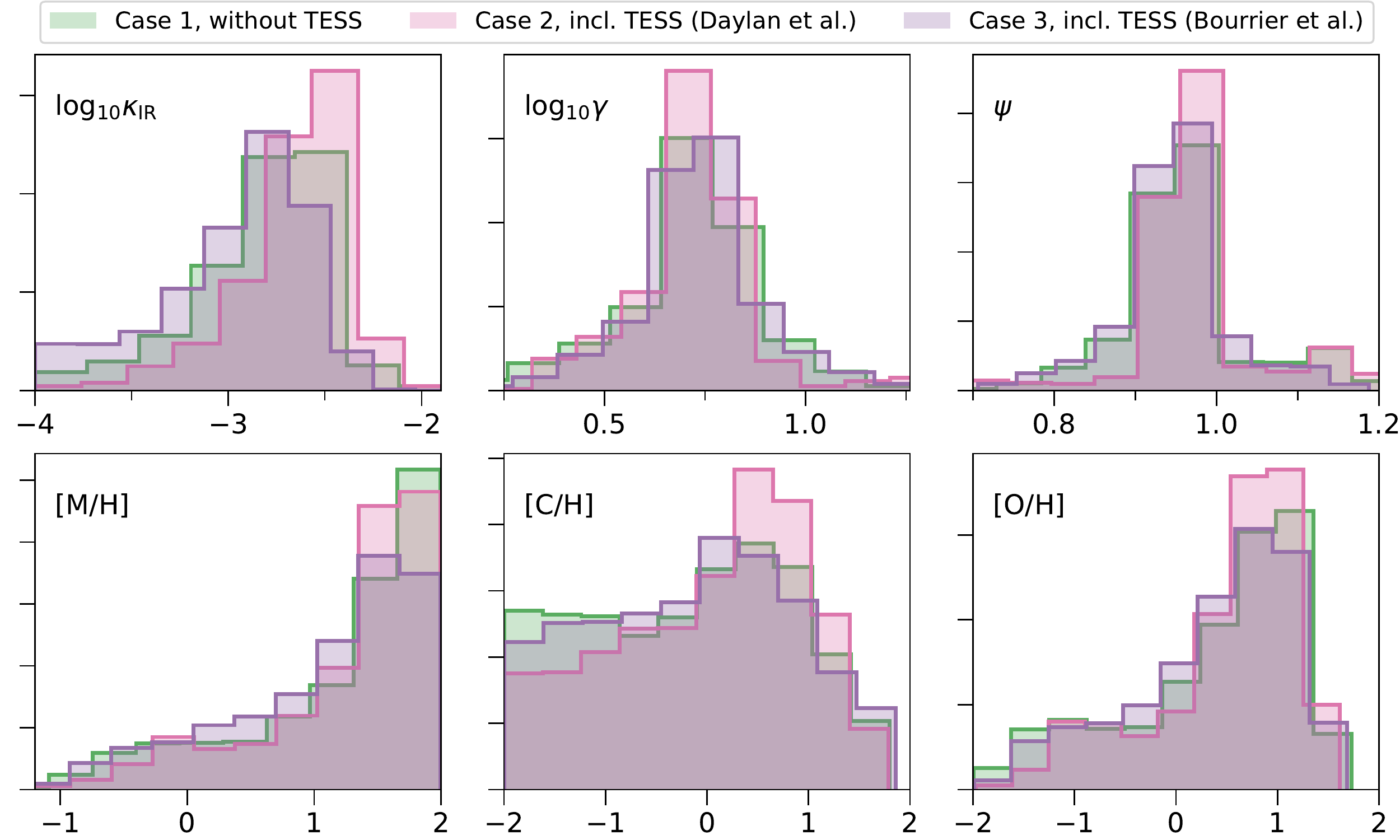}
\caption{Marginalized posterior distributions for the retrieval analyses described in the main text, which all adopt the updated WFC3 G141 emission spectrum presented in this work. Good agreement is obtained for all three cases. Note the lower bound on $\Cabun$ remains poorly constrained, as the available data do not spectrally resolve any carbon-bearing species. The upper bound of $\Z$ also remains unconstrained, due to the imposition of a hard limit of $<2$\,dex in the retrievals reported here, but future analyses should consider relaxing this assumption.}
\label{fig:posteriors}
\end{figure*}

\begin{table*}
\begin{minipage}{\linewidth}
  \centering
%\scriptsize
\caption{Retrieval uniform prior ranges and MCMC marginalized posterior distribution medians and $\pm$34\% credible intervals \label{table:posteriors}}
\begin{tabular}{ccccccc}
% \hline \\ 
\hline           &      &               & & \multicolumn{3}{c}{Fitting to updated WFC3 G141$^{\dagger}$}  \\ \cline{5-7}
 Parameter & Unit & Allowed range$^{\S}$ & \cite{2019MNRAS.488.2222M} & Case 1 & Case 2 & Case 3$^{\star}$  \medskip \\ \cline{1-7}
 &&& \\
 $\Z$ & dex &                              $-2$ to $2$ & ${1.09}_{-0.69}^{+0.57}$    & $1.57_{-0.94}^{+0.30}$ & $1.38_{-1.12}^{+0.42}$ & $1.50_{-0.75}^{+0.31}$ \smallskip \\
 $\Cabun$ & dex &                          $-2$ to $2$ & ${-0.29}_{-0.48}^{+0.61}$   & $0.05_{-1.38}^{+0.96}$ & $-0.13_{-1.20}^{+0.85}$ & $0.29_{-1.29}^{+0.70}$  \smallskip \\
 $\Oabun$ & dex &                          $-2$ to $2$ & ${0.18}_{-0.60}^{+0.64}$    & $0.78_{-1.22}^{+0.44}$ & $0.56_{-1.03}^{+0.49}$ & $0.71_{-0.72}^{+0.41}$  \smallskip \\
 $\log_{10}(\kir)$ & dex\,cm$^2$\,g$^{-1}$ & $-5$ to $0.5$ & ${-3.01}_{-0.62}^{+0.56}$ & $-2.70_{-0.36}^{+0.22}$ & $-2.85_{-0.55}^{+0.27}$ & $-2.61_{-0.31}^{+0.19}$ \smallskip \\
 $\log_{10}(\gamma)$ & dex &                $-4$ to $1.5$ & ${0.64}_{-0.16}^{+0.19}$  & $0.73_{-0.14}^{+0.12}$ & $0.74_{-0.14}^{+0.15}$ & $0.73_{-0.10}^{+0.10}$ \smallskip \\
 $\psi$ & ---  &                            $0$ to $2$   & ${0.99}_{-0.09}^{+0.06}$  & $0.95_{-0.05}^{+0.04}$ & $0.95_{-0.06}^{+0.05}$ & $0.97_{-0.03}^{+0.03}$ \medskip \\ \cline{1-7} 
\multicolumn{3}{r}{Best-fit model reduced $\chi^2$} & 0.85 & 0.78 & 0.94 & 0.79  \\ \hline  
\end{tabular}
\vspace{5pt} \\
\raggedright
$^{\dagger}$\, \footnotesize The three cases are those described in the main text: (1) fitting to the same dataset as \cite{2019MNRAS.488.2222M}, but using the updated WFC3 G141 spectrum derived in this work; (2) also including the \cite{2019arXiv190903000D} \textit{TESS} measurement; and (3) adopting the \cite{2019arXiv190903010B} \textit{TESS} measurement instead. \\
$^{\S}$\, \footnotesize Note that the allowed range for each of $\Z$, $\Cabun$, and $\Oabun$ was $-1$ to $2$ in \cite{2019MNRAS.488.2222M}. \\
$^{\star}$\, \footnotesize Our favored analysis, owing to it providing the tightest model parameter constraints and achieving the best fit quality (as quantified by the reduced $\chi^2$), while including all available data.
\end{minipage}
\end{table*}

The updated WASP-121b emission spectrum is shown in Figure \ref{fig:emspec}. The new G141 data is in good overall agreement with the original spectrum presented in \cite{2017Natur.548...58E}. However, unlike the \cite{2017Natur.548...58E} spectrum, the revised G141 spectrum does not exhibit a bump at $1.25\um$, which our previous investigations had failed to replicate with physically-plausible atmosphere models \citep{2017Natur.548...58E,2019MNRAS.488.2222M}. This suggests the $1.25\um$ bump was either a statistical fluctuation or a systematic artefact specific to the G141v0ec0 dataset, demonstrating the benefit gained by observing multiple eclipses well separated in time. Furthermore, the median eclipse depth uncertainty across the spectroscopic channels has improved from 90\,ppm \citep{2017Natur.548...58E} to 60\,ppm, bringing the H$_2$O emission band into sharper focus across the $\sim 1.3$-$1.6\,\um$ wavelength range. Despite the smaller uncertainties, the updated emission spectrum agrees with the best-fit retrieval model presented in \cite{2019MNRAS.488.2222M} (hereafter, ME19), reproduced in Figure \ref{fig:emspec}, which assumes equilibrium chemistry and accounts for the effects of thermal ionization and dissociation of molecules. This model has a temperature inversion, departing from a blackbody spectrum shortward of $\sim 1.3\,\um$ due to H$^{-}$ bound-free emission, between $\sim 1.3$-$1.6\,\um$ due to H$_2$O emission, and within the $4.5\um$ IRAC passband due to CO emission. Remarkably, with the revised WFC3 G141 spectrum, the $\chi^2$ value has improved from $43.6$ to $35.5$ for $42$ degrees of freedom, without any further tuning of the model. In Section \ref{sec:discussion:retrieval} below, we describe updated retrieval analyses performed on the revised dataset.

Phase curve measurements for WASP-121b made using TESS have also recently been reported by \cite{2019arXiv190903010B} (B19) and \cite{2019arXiv190903000D} (D19), covering the $0.6$-$1\,\um$ red optical wavelength range. For the planet-to-star dayside emission in the TESS passband, B19 obtain $419_{-42}^{+47}$ ppm, which agrees with the ME19 best-fit retrieval model shown in Figure \ref{fig:emspec} at the $0.4\sigma$ level. D19 report a somewhat higher dayside emission value of $534_{-43}^{+42}$ ppm, which is $2.3\sigma$ above the prediction of the best-fit ME19 model. However, D19 used the same retrieval methodology as ME19 and presented a best-fit model that is consistent with their TESS data point at the $0.7\sigma$ level. Given the posterior distributions of the ME19 and D19 retrieval analyses are consistent to within $1\sigma$ for all free parameters (i.e. $\Cabun$, $\Oabun$, $\Z$, $\kir$, $\gamma$, $\psi$), we deduce that the difference between the two available TESS analyses does not significantly affect the overall interpretation of the WASP-121b dayside spectrum. Although the best-fit models vary slightly depending on which TESS analysis is adopted, the posterior distributions are affected minimally and the conclusion that the dayside atmosphere of WASP-121b has a thermal inversion remains unchanged. This is explored further in Section \ref{sec:discussion:retrieval}.

Finally, if we assume the planet radiates as an isothermal blackbody and adopt the B19 TESS data point,\footnote{Note that for these calculations, we use a planet-to-star radius ratio of $\RpRs=0.1205$, approximately corresponding to the lowest point of the \cite{2018AJ....156..283E} transmission spectrum.} we obtain a best-fit temperature of $2703 \pm 6$\,K. However, the fit to the data is poor (Figure \ref{fig:emspec}) and can be ruled out at $6.6\sigma$ confidence. If instead we adopt the D19 TESS data point, the best-fit temperature is indistinguishable ($2704 \pm 6$\,K) and can be ruled out at $7.6\sigma$ confidence. The updated G141 dataset presented here and the TESS measurements recently reported in the literature therefore reinforce the conclusion that the dayside emission of WASP-121b is strongly inconsistent with an isothermal blackbody and is instead well explained by an atmosphere model including a thermal inversion.

\subsection{Retrieval analyses} \label{sec:discussion:retrieval}

Using the updated WFC3 G141 emission spectrum, we repeated the retrieval analysis described in ME19 for three separate dataset combinations: (Case 1) the WFC3 G102 and G141 spectrophotometry, \textit{Spitzer} IRAC photometry \citep{2020AJ....159..137G}, and published ground-based photometry \citep{2016MNRAS.tmp..312D,2019A&A...625A..80K}; (Case 2) the same, but also including the \cite{2019arXiv190903000D} \textit{TESS} eclipse measurement; and (Case 3) the same again, but instead adopting the \cite{2019arXiv190903010B} \textit{TESS} measurement. Our retrieval framework utilizes the \texttt{ATMO} code of \cite{2015ApJ...804L..17T}, which has been further developed by \cite{2016ApJ...817L..19T,2017ApJ...841...30T,2017ApJ...850...46T,2019ApJ...876..144T}, \cite{2014A&A...564A..59A}, \cite{2016A&A...594A..69D}, \cite{2018MNRAS.474.5158G,2019MNRAS.482.4503G}, and \cite{2020arXiv200313717P}, and employed in numerous other exoplanet retrieval analyses \citep[e.g.][]{2017Natur.548...58E,2018AJ....156..283E,2018AJ....156..298A,2018Natur.557..526N,2018MNRAS.474.1705N,2020MNRAS.tmp.1223C,2017Sci...356..628W,2018AJ....155...29W,2020MNRAS.tmp.1223C}. As in ME19, the free parameters of our model were: the carbon abundance ($\Cabun$); oxygen abundance ($\Oabun$); metallicity of all other heavy elements ($\Z$); infrared opacity ($\kir$); ratio of the visible-to-infrared opacity ($\gamma=\kappa_{\textnormal{V}}/\kir$); and an irradiation efficiency factor ($\psi$). For additional details, refer to ME19.

\begin{figure}
\centering  % this centres figure in column
\includegraphics[width=\linewidth]{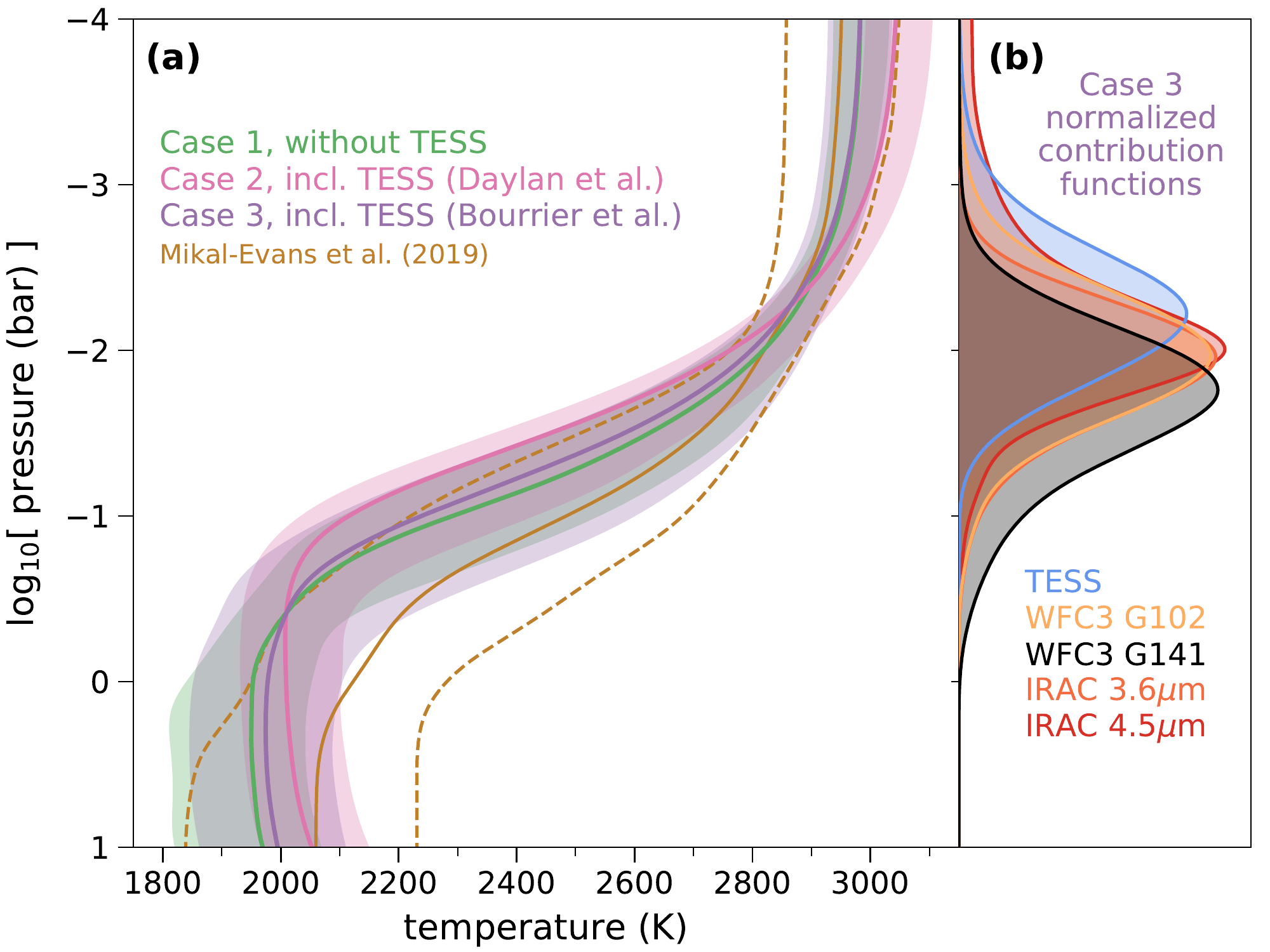}
\caption{(a) Retrieved PT profiles obtained for the three retrieval cases described in the main text. Solid lines indicate the median temperature at each pressure level across all PT profiles sampled by the MCMC analyses. Shaded regions indicate the temperature ranges at each pressure level encompassing $\pm$34\% of MCMC samples about the median. The corresponding PT distribution obtained in \citet{2019MNRAS.488.2222M} is also shown for comparison. (b) Normalized contribution functions corresponding to the best-fit model shown in Figure \ref{fig:emspec} for the Case 3 retrieval (i.e.\ including all latest available data and adopting the \citealp{2019arXiv190903010B} \textit{TESS} data point).}
\label{fig:pt}
\end{figure}

The best-fit spectra obtained for each retrieval are all in close agreement and plotted in Figure \ref{fig:emspec}. The marginalized posterior distributions for the model parameters are reported in Table \ref{table:posteriors} and shown in Figure \ref{fig:posteriors}. All three retrievals performed using the updated WFC3 G141 spectrum improve the constraints on the parameters controlling the PT profile (i.e.\ $\kir$, $\gamma$, $\psi$). This can be appreciated in Figure \ref{fig:pt}, which shows the PT profile distributions obtained for each retrieval analysis, along with the normalized contribution functions of the \textit{TESS}, \textit{HST}, and \textit{Spitzer} passbands.

For the parameters controlling elemental abundances (i.e. $\Z$, $\Cabun$, $\Oabun$), the results are also consistent with those reported in ME19. However, in the present work, we allowed the abundances to vary between $-2$ and 2 dex, whereas the lower bounds were set to $-1$ dex in ME19. Consequently, the posterior distributions we obtain here for $\Z$, $\Cabun$, and $\Oabun$ are typically broader than those reported in ME19. Despite this, for $\Z$ and $\Oabun$ the upper bounds are better constrained for each of the three retrievals performed in this work (Table \ref{table:posteriors}). This is likely due to two main reasons. First, the WFC3 G141 spectrum is dominated by an H$_2$O band, but does not encompass any strong bands due to carbon-based species. Hence, improving the precision on the WFC3 G141 spectrum results in a better constraint for $\Oabun$ while providing little additional information for $\Cabun$. Second, the better constrained H$_2$O abundance provides a reference level for the H$^-$ bound-free continuum, which spans the WFC3 G102 passband and short-wavelength half of the WFC3 G141 passband (e.g.\ see Figure 1 of \citealp{2018ApJ...855L..30A} and Figure 12 of ME19). This serves to calibrate the H$^{-}$ abundance, and hence the free electron abundance of the atmosphere, which is closely linked to $\Z$ via ionized heavy elements such as Na and K.

We also note that our retrieval results are overall consistent with those presented in D19. This is to be expected, as identical retrieval methodologies were employed in both studies and the same data were analyzed, with the exception of the updated WFC3 G141 spectrum adopted in this work. Additionally, our retrieval analyses are complementary to that presented in B19. In particular, the B19 study allowed chemical abundances to vary freely, whereas our retrieval analyses enforced chemical equilibrium while allowing $\Z$, $\Cabun$, and $\Oabun$ to vary. Despite these differences, visual inspection suggests the retrieved PT profile of B19 is in good agreement with those shown in Figure \ref{fig:pt}, increasing from $\sim 2200$K at 100\,mbar to $\sim 2900$\,K at 10\,mbar.

\section{Conclusion} \label{sec:conclusion}

We presented four new secondary eclipse observations of WASP-121b made with HST/WFC3 using the G141 grism, adding to the single eclipse observation previously reported in \cite{2017Natur.548...58E}. The additional data significantly increases the signal-to-noise of the measured dayside emission spectrum, with the median eclipse depth uncertainty reducing from 90ppm to 60ppm in 28 spectroscopic channels spanning the $1.12$-$1.64\,\um$ wavelength range. The updated spectrum is in excellent agreement with the best-fit model presented in \cite{2019MNRAS.488.2222M}, exhibiting an H$_2$O emission feature in the G141 passband, muted in amplitude due to thermal dissociation. Retrieval analyses performed using the updated WFC3 G141 spectrum allow tighter constraints to be placed on the PT profile in particular. These results reinforce the conclusion of previous studies \citep{2017Natur.548...58E,2019MNRAS.488.2222M,2019arXiv190903010B,2019arXiv190903000D} that the dayside hemisphere of WASP-121b has a thermal inversion.

\section*{Acknowledgements}

The authors are grateful to the anonymous referee for their constructive feedback. Based on observations made with the NASA/ESA Hubble Space Telescope, obtained from the data archive at the Space Telescope Science Institute. STScI is operated by the Association of Universities for Research in Astronomy, Inc. under NASA contract NAS 5-26555. Support for this work was provided by NASA through grant number GO-15134 from the Space Telescope Science Institute, which is operated by AURA, Inc., under NASA contract NAS 5-26555.

%\clearpage 
\bibliographystyle{apj}
\bibliography{wasp121}

\begin{thebibliography}{}
\expandafter\ifx\csname natexlab\endcsname\relax\def\natexlab#1{#1}\fi

\bibitem[{{Alam} {et~al.}(2018){Alam}, {Nikolov}, {L{\'o}pez-Morales}, {Sing},
  {Goyal}, {Henry}, {Sanz-Forcada}, {Williamson}, {Evans}, {Wakeford}, {Bruno},
  {Ballester}, {Stevenson}, {Lewis}, {Barstow}, {Bourrier}, {Buchhave},
  {Ehrenreich}, \& {Garc{\'{\i}}a Mu{\~n}oz}}]{2018AJ....156..298A}
{Alam}, M.~K., {Nikolov}, N., {L{\'o}pez-Morales}, M., {et~al.} 2018, \aj, 156,
  298

\bibitem[{{Amundsen} {et~al.}(2014){Amundsen}, {Baraffe}, {Tremblin},
  {Manners}, {Hayek}, {Mayne}, \& {Acreman}}]{2014A&A...564A..59A}
{Amundsen}, D.~S., {Baraffe}, I., {Tremblin}, P., {et~al.} 2014, \aap, 564, A59

\bibitem[{{Arcangeli} {et~al.}(2018){Arcangeli}, {D{\'e}sert}, {Line}, {Bean},
  {Parmentier}, {Stevenson}, {Kreidberg}, {Fortney}, {Mansfield}, \&
  {Showman}}]{2018ApJ...855L..30A}
{Arcangeli}, J., {D{\'e}sert}, J.-M., {Line}, M.~R., {et~al.} 2018, \apjl, 855,
  L30

\bibitem[{{Bourrier} {et~al.}(2019){Bourrier}, {Kitzmann}, {Kuntzer},
  {Nascimbeni}, {Lendl}, {Lavie}, {Hoeijmakers}, {Pino}, {Ehrenreich}, {Heng},
  {Allart}, {Cegla}, {Dumusque}, {Melo}, {Astudillo-Defru}, {Caldwell},
  {Cretignier}, {Giles}, {Henze}, {Jenkins}, {Lovis}, {Murgas}, {Pepe},
  {Ricker}, {Rose}, {Seager}, {Segransan}, {Suarez-Mascareno}, {Udry},
  {Vanderspek}, \& {Wyttenbach}}]{2019arXiv190903010B}
{Bourrier}, V., {Kitzmann}, D., {Kuntzer}, T., {et~al.} 2019, arXiv e-prints,
  arXiv:1909.03010

\bibitem[{{Bourrier} {et~al.}(2020){Bourrier}, {Ehrenreich}, {Lendl},
  {Cretignier}, {Allart}, {Dumusque}, {Cegla}, {Suarez-Mascareno},
  {Wyttenbach}, {Hoeijmakers}, {Melo}, {Kuntzer}, {Astudillo-Defru}, {Giles},
  {Heng}, {Kitzmann}, {Lavie}, {Lovis}, {Murgas}, {Nascimbeni}, {Pepe}, {Pino},
  {Segransan}, \& {Udry}}]{2020arXiv200106836B}
{Bourrier}, V., {Ehrenreich}, D., {Lendl}, M., {et~al.} 2020, arXiv e-prints,
  arXiv:2001.06836

\bibitem[{{Cabot} {et~al.}(2020){Cabot}, {Madhusudhan}, {Welbanks}, {Piette},
  \& {Gandhi}}]{2020arXiv200107196C}
{Cabot}, S. H.~C., {Madhusudhan}, N., {Welbanks}, L., {Piette}, A., \&
  {Gandhi}, S. 2020, arXiv e-prints, arXiv:2001.07196

\bibitem[{{Carter} {et~al.}(2020){Carter}, {Nikolov}, {Sing}, {Alam}, {Goyal},
  {Mikal-Evans}, {Wakeford}, {Henry}, {Morrell}, {L{\'o}pez-Morales},
  {Smalley}, {Lavvas}, {Barstow}, {Mu{\~n}oz}, {Gibson}, \&
  {Wilson}}]{2020MNRAS.tmp.1223C}
{Carter}, A.~L., {Nikolov}, N., {Sing}, D.~K., {et~al.} 2020, \mnras,
  arXiv:1911.12628

\bibitem[{{Daylan} {et~al.}(2019){Daylan}, {G{\"u}nther}, {Mikal-Evans},
  {Sing}, {Wong}, {Shporer}, {Crossfield}, {Niraula}, {de Wit}, {Koll},
  {Parmentier}, {Fetherolf}, {Kane}, {Ricker}, {Vand erspek}, {Seager}, {Winn},
  {Jenkins}, {Caldwell}, {Charbonneau}, {Henze}, {Paegert}, {Rinehart}, {Rose},
  {Sha}, {Quintana}, \& {Villasenor}}]{2019arXiv190903000D}
{Daylan}, T., {G{\"u}nther}, M.~N., {Mikal-Evans}, T., {et~al.} 2019, arXiv
  e-prints, arXiv:1909.03000

\bibitem[{{Delrez} {et~al.}(2016){Delrez}, {Santerne}, {Almenara}, {Anderson},
  {Collier-Cameron}, {D{\'{\i}}az}, {Gillon}, {Hellier}, {Jehin}, {Lendl},
  {Maxted}, {Neveu-VanMalle}, {Pepe}, {Pollacco}, {Queloz}, {S{\'e}gransan},
  {Smalley}, {Smith}, {Triaud}, {Udry}, {Van Grootel}, \&
  {West}}]{2016MNRAS.tmp..312D}
{Delrez}, L., {Santerne}, A., {Almenara}, J.-M., {et~al.} 2016, \mnras,
  doi:10.1093/mnras/stw522

\bibitem[{{Drummond} {et~al.}(2016){Drummond}, {Tremblin}, {Baraffe},
  {Amundsen}, {Mayne}, {Venot}, \& {Goyal}}]{2016A&A...594A..69D}
{Drummond}, B., {Tremblin}, P., {Baraffe}, I., {et~al.} 2016, \aap, 594, A69

\bibitem[{{Evans} {et~al.}(2016){Evans}, {Sing}, {Wakeford}, {Nikolov},
  {Ballester}, {Drummond}, {Kataria}, {Gibson}, {Amundsen}, \&
  {Spake}}]{2016ApJ...822L...4E}
{Evans}, T.~M., {Sing}, D.~K., {Wakeford}, H.~R., {et~al.} 2016, \apjl, 822, L4

\bibitem[{{Evans} {et~al.}(2017){Evans}, {Sing}, {Kataria}, {Goyal}, {Nikolov},
  {Wakeford}, {Deming}, {Marley}, {Amundsen}, {Ballester}, {Barstow},
  {Ben-Jaffel}, {Bourrier}, {Buchhave}, {Cohen}, {Ehrenreich}, {Garc{\'{\i}}a
  Mu{\~n}oz}, {Henry}, {Knutson}, {Lavvas}, {Lecavelier Des Etangs}, {Lewis},
  {L{\'o}pez-Morales}, {Mandell}, {Sanz-Forcada}, {Tremblin}, \&
  {Lupu}}]{2017Natur.548...58E}
{Evans}, T.~M., {Sing}, D.~K., {Kataria}, T., {et~al.} 2017, \nat, 548, 58

\bibitem[{{Evans} {et~al.}(2018){Evans}, {Sing}, {Goyal}, {Nikolov}, {Marley},
  {Zahnle}, {Henry}, {Barstow}, {Alam}, {Sanz-Forcada}, {Kataria}, {Lewis},
  {Lavvas}, {Ballester}, {Ben-Jaffel}, {Blumenthal}, {Bourrier}, {Drummond},
  {Garc{\'{\i}}a Mu{\~n}oz}, {L{\'o}pez-Morales}, {Tremblin}, {Ehrenreich},
  {Wakeford}, {Buchhave}, {Lecavelier des Etangs}, {H{\'e}brard}, \&
  {Williamson}}]{2018AJ....156..283E}
{Evans}, T.~M., {Sing}, D.~K., {Goyal}, J.~M., {et~al.} 2018, \aj, 156, 283

\bibitem[{{Fortney} {et~al.}(2008){Fortney}, {Lodders}, {Marley}, \&
  {Freedman}}]{2008ApJ...678.1419F}
{Fortney}, J.~J., {Lodders}, K., {Marley}, M.~S., \& {Freedman}, R.~S. 2008,
  \apj, 678, 1419

\bibitem[{{Gandhi} \& {Madhusudhan}(2019)}]{2019MNRAS.485.5817G}
{Gandhi}, S., \& {Madhusudhan}, N. 2019, \mnras, 485, 5817

\bibitem[{{Garhart} {et~al.}(2020){Garhart}, {Deming}, {Mandell}, {Knutson},
  {Wallack}, {Burrows}, {Fortney}, {Hood}, {Seay}, {Sing}, {Benneke}, {Fraine},
  {Kataria}, {Lewis}, {Madhusudhan}, {McCullough}, {Stevenson}, \&
  {Wakeford}}]{2020AJ....159..137G}
{Garhart}, E., {Deming}, D., {Mandell}, A., {et~al.} 2020, \aj, 159, 137

\bibitem[{{Gibson} {et~al.}(2020){Gibson}, {Merritt}, {Nugroho}, {Cubillos},
  {de Mooij}, {Mikal-Evans}, {Fossati}, {Lothringer}, {Nikolov}, {Sing},
  {Spake}, {Watson}, \& {Wilson}}]{2020MNRAS.tmp..220G}
{Gibson}, N.~P., {Merritt}, S., {Nugroho}, S.~K., {et~al.} 2020, \mnras, 220

\bibitem[{{Goyal} {et~al.}(2019){Goyal}, {Wakeford}, {Mayne}, {Lewis},
  {Drummond}, \& {Sing}}]{2019MNRAS.482.4503G}
{Goyal}, J.~M., {Wakeford}, H.~R., {Mayne}, N.~J., {et~al.} 2019, \mnras, 482,
  4503

\bibitem[{{Goyal} {et~al.}(2018){Goyal}, {Mayne}, {Sing}, {Drummond},
  {Tremblin}, {Amundsen}, {Evans}, {Carter}, {Spake}, {Baraffe}, {Nikolov},
  {Manners}, {Chabrier}, \& {Hebrard}}]{2018MNRAS.474.5158G}
{Goyal}, J.~M., {Mayne}, N., {Sing}, D.~K., {et~al.} 2018, \mnras, 474, 5158

\bibitem[{{Hubeny} {et~al.}(2003){Hubeny}, {Burrows}, \&
  {Sudarsky}}]{2003ApJ...594.1011H}
{Hubeny}, I., {Burrows}, A., \& {Sudarsky}, D. 2003, \apj, 594, 1011

\bibitem[{{Knutson} {et~al.}(2014){Knutson}, {Benneke}, {Deming}, \&
  {Homeier}}]{2014Natur.505...66K}
{Knutson}, H.~A., {Benneke}, B., {Deming}, D., \& {Homeier}, D. 2014, \nat,
  505, 66

\bibitem[{{Kov{\'a}cs} \& {Kov{\'a}cs}(2019)}]{2019A&A...625A..80K}
{Kov{\'a}cs}, G., \& {Kov{\'a}cs}, T. 2019, \aap, 625, A80

\bibitem[{{Kreidberg} {et~al.}(2014){Kreidberg}, {Bean}, {D{\'e}sert},
  {Benneke}, {Deming}, {Stevenson}, {Seager}, {Berta-Thompson}, {Seifahrt}, \&
  {Homeier}}]{2014Natur.505...69K}
{Kreidberg}, L., {Bean}, J.~L., {D{\'e}sert}, J.-M., {et~al.} 2014, \nat, 505,
  69

\bibitem[{{Lothringer} {et~al.}(2018){Lothringer}, {Barman}, \&
  {Koskinen}}]{2018ApJ...866...27L}
{Lothringer}, J.~D., {Barman}, T., \& {Koskinen}, T. 2018, \apj, 866, 27

\bibitem[{{Mikal-Evans} {et~al.}(2019){Mikal-Evans}, {Sing}, {Goyal},
  {Drummond}, {Carter}, {Henry}, {Wakeford}, {Lewis}, {Marley}, {Tremblin},
  {Nikolov}, {Kataria}, {Deming}, \& {Ballester}}]{2019MNRAS.488.2222M}
{Mikal-Evans}, T., {Sing}, D.~K., {Goyal}, J.~M., {et~al.} 2019, \mnras, 488,
  2222

\bibitem[{{Nikolov} {et~al.}(2018{\natexlab{a}}){Nikolov}, {Sing}, {Fortney},
  {Goyal}, {Drummond}, {Evans}, {Gibson}, {De Mooij}, {Rustamkulov},
  {Wakeford}, {Smalley}, {Burgasser}, {Hellier}, {Helling}, {Mayne},
  {Madhusudhan}, {Kataria}, {Baines}, {Carter}, {Ballester}, {Barstow},
  {McCleery}, \& {Spake}}]{2018Natur.557..526N}
{Nikolov}, N., {Sing}, D.~K., {Fortney}, J.~J., {et~al.} 2018{\natexlab{a}},
  \nat, 557, 526

\bibitem[{{Nikolov} {et~al.}(2018{\natexlab{b}}){Nikolov}, {Sing}, {Goyal},
  {Henry}, {Wakeford}, {Evans}, {L{\'o}pez-Morales}, {Garc{\'{\i}}a Mu{\~n}oz},
  {Ben-Jaffel}, {Sanz-Forcada}, {Ballester}, {Kataria}, {Barstow}, {Bourrier},
  {Buchhave}, {Cohen}, {Deming}, {Ehrenreich}, {Knutson}, {Lavvas}, {Lecavelier
  des Etangs}, {Lewis}, {Mandell}, \& {Williamson}}]{2018MNRAS.474.1705N}
{Nikolov}, N., {Sing}, D.~K., {Goyal}, J., {et~al.} 2018{\natexlab{b}}, \mnras,
  474, 1705

\bibitem[{{Parmentier} {et~al.}(2018){Parmentier}, {Line}, {Bean}, {Mansfield},
  {Kreidberg}, {Lupu}, {Visscher}, {D{\'e}sert}, {Fortney}, {Deleuil},
  {Arcangeli}, {Showman}, \& {Marley}}]{2018A&A...617A.110P}
{Parmentier}, V., {Line}, M.~R., {Bean}, J.~L., {et~al.} 2018, \aap, 617, A110

\bibitem[{{Phillips} {et~al.}(2020){Phillips}, {Tremblin}, {Baraffe},
  {Chabrier}, {Allard}, {Spiegelman}, {Goyal}, {Drummond}, \&
  {Hebrard}}]{2020arXiv200313717P}
{Phillips}, M.~W., {Tremblin}, P., {Baraffe}, I., {et~al.} 2020, arXiv
  e-prints, arXiv:2003.13717

\bibitem[{{Sing} {et~al.}(2019){Sing}, {Lavvas}, {Ballester}, {Lecavelier des
  Etangs}, {Marley}, {Nikolov}, {Ben-Jaffel}, {Bourrier}, {Buchhave}, {Deming},
  {Ehrenreich}, {Mikal-Evans}, {Kataria}, {Lewis}, {L{\'o}pez-Morales},
  {Garc{\'\i}a Mu{\~n}oz}, {Henry}, {Sanz-Forcada}, {Spake}, {Wakeford}, \&
  {PanCET Collaboration}}]{2019AJ....158...91S}
{Sing}, D.~K., {Lavvas}, P., {Ballester}, G.~E., {et~al.} 2019, \aj, 158, 91

\bibitem[{{Stevenson} \& {Fowler}(2019)}]{2019wfc..rept...12S}
{Stevenson}, K.~B., \& {Fowler}, J. 2019, {Analyzing Eight Years of Transiting
  Exoplanet Observations Using WFC3's Spatial Scan Monitor}, Tech. rep.

\bibitem[{{Tremblin} {et~al.}(2016){Tremblin}, {Amundsen}, {Chabrier},
  {Baraffe}, {Drummond}, {Hinkley}, {Mourier}, \&
  {Venot}}]{2016ApJ...817L..19T}
{Tremblin}, P., {Amundsen}, D.~S., {Chabrier}, G., {et~al.} 2016, \apjl, 817,
  L19

\bibitem[{{Tremblin} {et~al.}(2015){Tremblin}, {Amundsen}, {Mourier},
  {Baraffe}, {Chabrier}, {Drummond}, {Homeier}, \&
  {Venot}}]{2015ApJ...804L..17T}
{Tremblin}, P., {Amundsen}, D.~S., {Mourier}, P., {et~al.} 2015, \apjl, 804,
  L17

\bibitem[{{Tremblin} {et~al.}(2017{\natexlab{a}}){Tremblin}, {Chabrier},
  {Mayne}, {Amundsen}, {Baraffe}, {Debras}, {Drummond}, {Manners}, \&
  {Fromang}}]{2017ApJ...841...30T}
{Tremblin}, P., {Chabrier}, G., {Mayne}, N.~J., {et~al.} 2017{\natexlab{a}},
  \apj, 841, 30

\bibitem[{{Tremblin} {et~al.}(2017{\natexlab{b}}){Tremblin}, {Chabrier},
  {Baraffe}, {Liu}, {Magnier}, {Lagage}, {Alves de Oliveira}, {Burgasser},
  {Amundsen}, \& {Drummond}}]{2017ApJ...850...46T}
{Tremblin}, P., {Chabrier}, G., {Baraffe}, I., {et~al.} 2017{\natexlab{b}},
  \apj, 850, 46

\bibitem[{{Tremblin} {et~al.}(2019){Tremblin}, {Padioleau}, {Phillips},
  {Chabrier}, {Baraffe}, {Fromang}, {Audit}, {Bloch}, {Burgasser}, {Drummond},
  {Gonz{\'a}lez}, {Kestener}, {Kokh}, {Lagage}, \&
  {Stauffert}}]{2019ApJ...876..144T}
{Tremblin}, P., {Padioleau}, T., {Phillips}, M.~W., {et~al.} 2019, \apj, 876,
  144

\bibitem[{{Wakeford} {et~al.}(2017){Wakeford}, {Sing}, {Kataria}, {Deming},
  {Nikolov}, {Lopez}, {Tremblin}, {Amundsen}, {Lewis}, {Mandell}, {Fortney},
  {Knutson}, {Benneke}, \& {Evans}}]{2017Sci...356..628W}
{Wakeford}, H.~R., {Sing}, D.~K., {Kataria}, T., {et~al.} 2017, Science, 356,
  628

\bibitem[{{Wakeford} {et~al.}(2018){Wakeford}, {Sing}, {Deming}, {Lewis},
  {Goyal}, {Wilson}, {Barstow}, {Kataria}, {Drummond}, {Evans}, {Carter},
  {Nikolov}, {Knutson}, {Ballester}, \& {Mandell}}]{2018AJ....155...29W}
{Wakeford}, H.~R., {Sing}, D.~K., {Deming}, D., {et~al.} 2018, \aj, 155, 29

\end{thebibliography}

\end{document}